\documentclass[conference]{IEEEtran}
\IEEEoverridecommandlockouts
\usepackage{cite}
\usepackage{amsmath,amssymb,amsfonts}
\usepackage{algorithmic}
\usepackage{graphicx}
\usepackage{textcomp}
\usepackage{xcolor}
\def\BibTeX{{\rm B\kern-.05em{\sc i\kern-.025em b}\kern-.08em
    T\kern-.1667em\lower.7ex\hbox{E}\kern-.125emX}}
\usepackage[utf8]{inputenc}
\usepackage[margin=0.75in]{geometry}
\usepackage{hyperref}
\usepackage{graphicx}
\usepackage{xspace}

\usepackage{listings}

\newcommand{\uclid}{Uclid5\xspace}
\newcommand{\pmpchecker}{\texttt{PMPChecker}\xspace}
\title{Verifying RISC-V Physical Memory Protection}

\author{
\IEEEauthorblockN{Kevin Cheang, Cameron Rasmussen, Dayeol Lee, David W. Kohlbrenner, Krste Asanovi\'c, Sanjit A. Seshia}
\IEEEauthorblockA{\textit{Department of Electrical Engineering and Computer Sciences} \\
\textit{University of California, Berkeley}\\
Email: \{kcheang, crasmussen, dayeol, dkohlbre, krste, sseshia\}@berkeley.edu
}

}

\begin{document}

\maketitle

\begin{abstract}
We formally verify an open-source hardware implementation of physical memory protection (PMP) in RISC-V, which is a standard feature used for memory isolation in security critical systems such as the Keystone trusted execution environment.
PMP provides per-hardware-thread machine-mode control registers that specify the access privileges for physical memory regions.
We first formalize the functional property of the PMP rules based on the RISC-V ISA manual. 
Then, we use the LIME tool to translate an open-source implementation of the PMP hardware module written in Chisel to the \uclid formal verification language.
We encode the formal specification in \uclid and verify the functional correctness of the hardware.
This is an initial effort towards verifying the Keystone framework, where the trusted computing base (TCB) relies on PMP to provide security guarantees such as integrity and confidentiality.
\end{abstract}

\section{Introduction}




Physical memory protection (PMP) is a standard RISC-V feature that allows the firmware to specify physical memory regions and control the memory access permissions.
Many systems have adopted PMP to protect memory regions for high-privilege binaries (e.g., firmware) or devices.
For example, OpenSBI~\cite{opensbi} uses PMP to allow the firmware to protect its own memory region when the machine boots.
PMP has been also used in trusted execution environments based on RISC-V~\cite{keystone,hexfive}.
Keystone~\cite{keystone} uses multiple PMP entries to isolate each enclave from the rest of the system including the privileged operating system, and also to manage shared memory regions.
Keystone utilizes several PMP rules such as whitelist-based or prioritized address matching to implement a flexible memory isolation scheme.
Thus, it is fair to say that the entire security guarantee of Keystone relies on the functional correctness of PMP.

One way to ensure functional correctness of hardware or software is to use formal methods.
Formal methods provide machine-assisted proofs for a given formalization and can be used to ensure specific properties hold on a system implementation.
We claim that the formal verification of PMP is needed as a first step to ensure security guarantees of systems such as Keystone.
However, we find that the implementation of PMP rules has not been previously formally verified.
We also find that the PMP rules are well-defined yet not formally specified.


In our work, we formally verify the hardware implementation of PMP rules.
First, we provide a formal specification of the PMP feature in RISC-V ISA. To model the hardware implementation, we automatically generate the formal model of the PMP module in an open-source RISC-V core, Rocket Chip~\cite{rocket}, by using a tool, LIME~\cite{goldengatelime}.
LIME can translate the FIRRTL~\cite{firrtl}, an intermediate representation of Chisel~\cite{chisel} hardware description language, to the \uclid~\cite{uclid5} verification language.
Then, we encode the specification of PMP based off of the RISC-V ISA manual to verify the functional correctness of the module, \texttt{PMPChecker}, a core unit for PMP.
We also verify a restricted configuration of the PMP unit to ease the verification effort and describe this in the evaluation section.

Our verification results show that the current implementation of PMP rules in Rocket Chip is functionally correct.
However, we acknowledge that it does not imply the functional correctness of the entire PMP implementation.
The correctness of PMP not only relies on other hardware components such as translation look-aside buffer (TLB) and page table walker (PTW), but also requires a correct software implementation.
For example, a memory access may bypass the PMP rule if the address is cached in TLB.
In order to prevent this, most systems including Keystone flush the TLB whenever it changes the local PMP policy.
Also, previous work~\cite{nelsonsm} has reported that 
a bug in other hardware component can cause a failure on
memory accesses which \pmpchecker allows.
We plan to extend our verification to include the other hardware components that may affect the actual PMP enforcement as well as the software that uses PMP.

\section{Related Work}
Existing commercial implementations of enclaves like Intel's SGX~\cite{sgx} lack transparency on formal correctness guarantees and higher level security properties such as confidentiality and integrity. On the other hand, non-commercial implementations of enclaves, such as MIT's Sanctum~\cite{sanctum}, lack formal reasoning at the hardware level. Despite prior work on verifying the design of these enclaves~\cite{pramod-enclave} at a higher abstraction level, there is little work on verifying the full implementation details of the underlying hardware at the RTL implementation level. This work aims to fill this gap by reasoning about PMP at the RTL level to provide strong security guarantees for enclaves built using Keystone and in general, for applications that use PMP.

\section{Background}

\subsection{Physical Memory Protection (PMP) and Keystone}
\begin{figure}
    \centering
    \includegraphics[width=.95\linewidth]{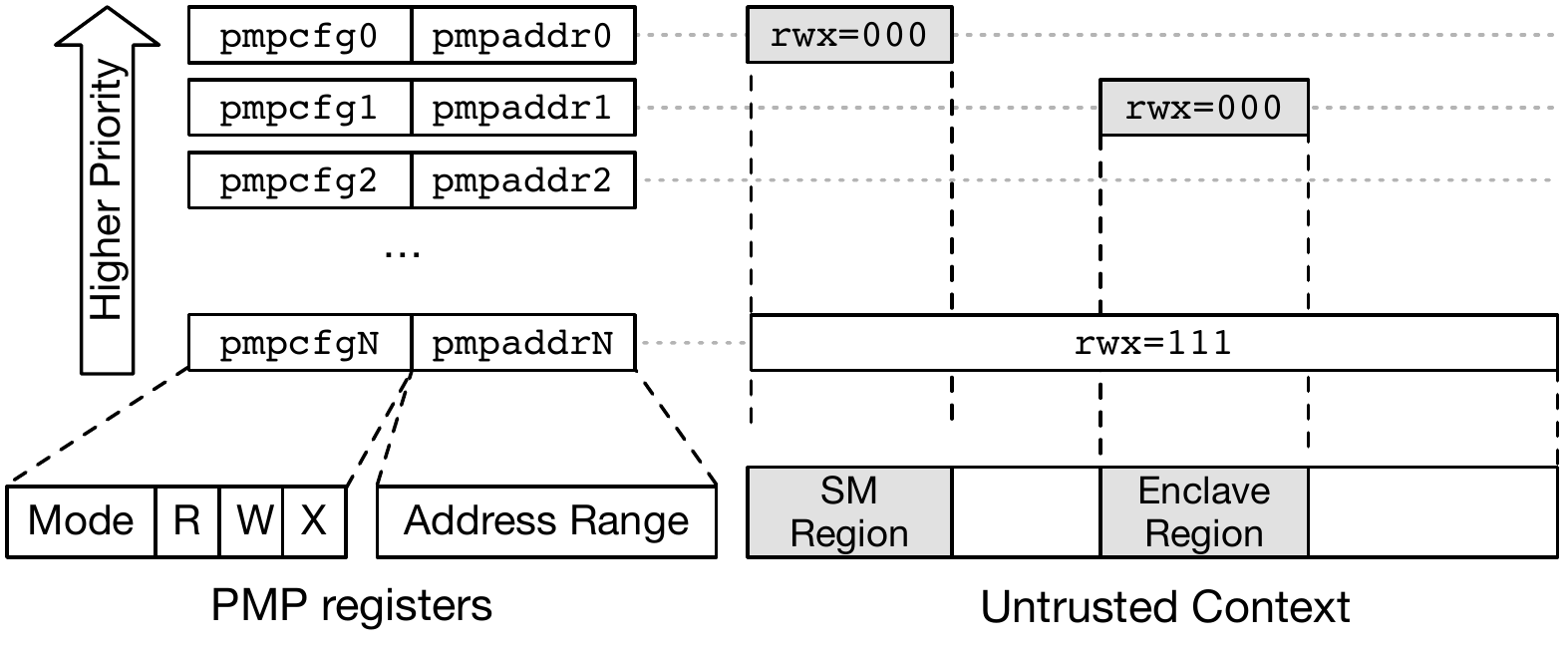}
    \caption{An example usage of RISC-V PMP: memory isolation in Keystone enclaves.}
    \label{fig:pmp}
\end{figure}

To provide an overview, PMP controls the access permissions to a specified physical memory region, by using a set of control status registers (CSR) in RISC-V.
Each core may have 0-16 PMP registers, each of which consists of a configuration (\texttt{pmpcfg}) and an address register (\texttt{pmpaddr}) to define a \textit{PMP entry}.
As shown in Figure \ref{fig:pmp}, the \texttt{pmpcfg} register defines the addressing mode and permission bits, and \texttt{pmpaddr} specifies the address range by encoding the address using a selected addressing mode.
There are three addressing modes: 4-byte aligned word (NA4), naturally-aligned power-of-two (NAPOT), or top-of-range (TOR).
PMP entries act as a whitelist, which means that the memory is inaccessible if none of the PMP entries are defined.
The PMP entries are also statically prioritized, such that the lowest-numbered PMP entry that matches any byte of a memory access in addition to the privilege mode, determines whether the memory access succeeds or fails~\cite{riscv-priv}.

These PMP rules are critical to the memory isolation in Keystone trusted execution environment (TEE).
When the system boots, a software called the security monitor in Keystone uses the first PMP register to protect its own memory region by setting all permission bits to zero and configuring the address to cover the entire image as well as the stack (Figure \ref{fig:pmp}).
Then, it sets the last (the lowest priority) PMP register to let the OS access the remaining part of the memory.
Upon the creation of an enclave, the security monitor allocates an available PMP register, to seal and isolate the enclave memory.
Because of the priority, the OS is never allowed to access either the security monitor's or enclave's memory.

In Keystone, the security monitor implements memory isolation by switching the permission bits when the context changes.
Before an enclave starts to run, the security monitor flips the permission bits in the enclave's PMP entry in order to allow computation on its isolated memory.
In addition, the security monitor invalidates the last PMP register, in order to deny the enclave access to the operating system's memory.

\section{Formal Specification}\label{label:spec}

PMP controls the memory access based on a few RISC-V control-and-status registers (CSRs).
PMP logically consists of multiple PMP \textit{entries} and each entry specifies a range of physical address and read, write, and execute permissions.
In Rocket Chip, a RISC-V open-source processor, PMP rules are implemented by a core hardware module called \pmpchecker.
We begin by defining the function of \pmpchecker and then define a set of primitive functions that abstractly describe the behavior of the \pmpchecker and finally state the functional property of the \pmpchecker. As a precursor, the \pmpchecker is a combinatorial logic circuit that takes in the address and the size of a memory access. However, it also takes inputs from the system, which are the PMP registers and the current privilege mode of the core.

First, we define the set of finite bit addresses to be $\mathcal{A}$, the set of PMP region indices to be $\mathcal{N}$, and the set of bitvectors of width $n$ to be $\{0,1\}^n$. Focusing on the \pmpchecker of the PMP unit, the set of argument variables of the \pmpchecker consists of the address to the \pmpchecker $\mathcal{I}_{addr} \in \{0,1\}^N$ (where N is the architecture address length), the size of the memory access $2^{\mathcal{I}_{size}}, \mathcal{I}_{size} \in \{0,1\}^2$, the current PMP register states $\mathcal{I}_{cfg}$ which is an array of type $cfg = \{ l, x, w, r \}$ (i.e., a struct of 1-bit variables: $l$ is the lock bit, and $r, w, x$ are the read, write, and execute permissions respectively), and the current privilege mode of the RISC-V system $\mathcal{I}_{prv}$ \footnote{To denote the $i^{th}$ PMP region's writable bit, we write $\mathcal{I}_{cfg}[i].w$.}. The output variables contain the permission of the memory access for the current privilege mode $\mathcal{I}_{prv}$, denoted by $\mathcal{O}_r, \mathcal{O}_w, \mathcal{O}_x$ as the read, write, and execute permissions respectively.

We now define the primitives used in our \pmpchecker property: let $r(addr, i): (addr, i) \mapsto bool$ be a function that returns true when the address $addr$ is contained within the $i^{th}$ PMP region and $a(addr, i): (addr, i)\mapsto bool$ be a function that returns true when the address $addr$ is within the region's mask (i.e., the address is aligned according to the addressing mode). To reason about whether an address $addr$ is within a region's boundary, we define $r_{lo}(i): i \mapsto \mathcal{A}$ and $r_{hi}(i): i \mapsto \mathcal{A}$ as functions that return the low and high address of the $i^{th}$ region. In the RISC-V ISA, $r_{lo}$ and $r_{hi}$ are defined by the addressing mode (e.g. NAPOT).

Then $r$ is defined as a function that returns true if and only if for a given address $addr$ and region $i$, address is between the respective low and high address boundaries of that region:
\begin{equation}
\begin{aligned}
    &\forall addr\in\mathcal{A}, \forall i \in \mathcal{N},\\
    &\qquad r(addr, i) \iff r_{lo}(i) \leq addr \leq r_{hi}(i)
\end{aligned}
\end{equation}
While $a$ is defined as a function that returns true when the given address is within the $i^{th}$ region's range and implies that the last byte accessed is also in bounds:
\begin{equation}
\begin{aligned}
    &\forall addr \in \mathcal{A},\forall size\in\{0,1\}^2, \forall i \in \mathcal{N},\\
    &\qquad(r(addr, i) \land a(addr, i)) \Rightarrow\\
    &\qquad\qquad(addr + (1 << size) - 1) \leq r_{hi}(i)
\end{aligned}
\end{equation}
The primary property of the \pmpchecker is that the returned permission bits correspond to the highest priority register that contains the queried address in its region with the following exceptions:
\begin{enumerate}
    \item If the address is not contained in any region, we return the default permissions
        \begin{itemize}
            \item High privilege modes - full permissions
            \item Low privilege modes - no permission
        \end{itemize}
    \item If we are operating in a high privilege mode
        \begin{itemize}
            \item If the region is not locked, then we have full permissions
            \item If the region is locked, then we only have access according to the PMP region's set permissions
        \end{itemize}
    \item If our access is large enough to only partially fall within the boundary of the highest priority region, it will deny all permissions
\end{enumerate}
We decompose this property into three separate properties.
First, let $low\in\{0,1\}$ represent the value that variable $prv$ evaluates to if it is in low privilege mode. Conversely, high privilege mode is represented by the negation of low: $high = \neg low$. Then the following first invariant captures the primary invariant without breaking the exceptions and accounts for exception 3 while the system operates in low privilege mode:
\begin{equation}
    \begin{aligned}
    &(\mathcal{I}_{prv} = low) \Rightarrow\\
    &\qquad\forall addr\in\mathcal{A}, \forall i\in\mathcal{N}, \label{eq:mainprop}\\
    & \qquad (r(addr, i) \land \neg(\exists j\in\mathcal{N}, j < i \land r(addr, j))) \Rightarrow\\
    & \qquad\qquad\mathcal{O}_{r} = (\mathcal{I}_{cfg}[i].r \land a(addr, i)) \land\\
    & \qquad\qquad\mathcal{O}_{w} = (\mathcal{I}_{cfg}[i].w \land a(addr, i)) \land\\
    & \qquad\qquad\mathcal{O}_{x} = (\mathcal{I}_{cfg}[i].x) \land a(addr, i))
    \end{aligned}
\end{equation}
To handle exception 1, where the address is not in any regions, we have the property:
\begin{equation}
    \begin{aligned}
    &\forall addr\in\mathcal{A}, \neg(\exists i\in\mathcal{N}, r(addr, i)) \Rightarrow \label{eq:ex1}\\
    &\qquad(\mathcal{O}_{r} = (\mathcal{I}_{prv} \neq low) \land\\
    &\qquad\mathcal{O}_{w} = (\mathcal{I}_{prv} \neq low) \land\\
    &\qquad\mathcal{O}_{x} = (\mathcal{I}_{prv} \neq low))
    \end{aligned}
\end{equation}
And finally for exception 2 and 3, when the system mode is in high privilege, we have the property:
\begin{equation}
    \begin{aligned}
    &(\mathcal{I}_{prv} \neq low) \Rightarrow\\
    &\qquad\forall addr\in\mathcal{A}, \forall i\in\mathcal{N}, \label{eq:ex23}\\
    &\qquad(r(addr, i) \land \neg(\exists j\in\mathcal{N}, j < i \land r(addr, j))) \Rightarrow\\
    &\qquad\qquad\mathcal{O}_{r} = ((\neg\mathcal{I}_{cfg}[i].l \lor \mathcal{I}_{cfg}[i].r)
    \land a(addr, i)) \land\\
    &\qquad\qquad\mathcal{O}_{w} = ((\neg\mathcal{I}_{cfg}[i].l \lor \mathcal{I}_{cfg}[i].w)
    \land a(addr, i)) \land\\
    &\qquad\qquad\mathcal{O}_{x} = ((\neg\mathcal{I}_{cfg}[i].l \lor \mathcal{I}_{cfg}[i].x)
    \land a(addr, i)))
    \end{aligned}
\end{equation}

\section{Evaluation}\label{label:eval}
For our evaluation, we focused on verifying the PMP FIRRTL implementation from the Rocket Chip core. More specifically, we verified the functional correctness of the \pmpchecker using the \uclid verification toolkit.

The scope of the verification effort is restricted to verifying the PMP model using the default configuration of the PMP implementation in Rocket Chip. We also only verify the \pmpchecker module of Rocket Chip, which is the core component that is queried and computes the permission bits on every memory access.


\subsection{Workflow}
To build our verification model, we first emit the FIRRTL implementation description for Rocket Chip by running a low-FIRRTL pass over the Chisel implementation using the Chisel generator. Then we use the LIME ~\cite{goldengatelime} translator to automatically translate the FIRRTL description to a \uclid model. We then extracted the \pmpchecker module from the \uclid model, specified, and verified the \pmpchecker model with our functional specification described in the previous section~\ref{label:spec} under specific default configurations \footnote{The models can be found at \url{https://github.com/veri-v/pmpcheckerspec}}.

\subsection{Results}
The Chisel implementation of the \pmpchecker contained 48 LoC which translated to \uclid models with 1125 LoC. When running \uclid on the model using Z3 as the backend solver, the engine completed the verification using 1-step induction in 41.331s (real time) on a 2.6 GHz Intel Core i7 machine with 16 GB RAM on OSX.

\section{Future Work}

Although we show functional correctness of the \texttt{PMPChecker} module, 
Rocket enforces PMP rules using multiple other hardware components including the translation look-side buffer (TLB) and page table walker (PTW).
When the core accesses an address, the TLB and the PTW will translate the virtual addresses into a physical address, and then \texttt{PMPChecker} return the permissions for that address given the core's current privilege mode.
If the access is restricted, the TLB entry for the address is prevented from being filled, such that the access raises access fault.
Thus, to verify the functional correctness of the entire PMP, we also need to verify the composition of these other components.
Also, the higher-level properties such as memory isolation will not only rely on the functional correctness of hardware, but also on the functional correctness of software interacting with hardware.
To this end, we are planning to formally verify other hardware components as well as the Keystone security monitor as an example software implementation.

\section{Conclusion}

To conclude, we have provided a formal specification of the \pmpchecker, which is a core component of the PMP feature in the RISC-V ISA. Using a Chisel generator and the LIME transpiler, we automatically generated an implementation accurate model of the \pmpchecker from an implementation of RISC-V and, Rocket Chip. We specified the functional properties of the \pmpchecker module and verified it using \uclid. This is a first step towards verifying Keystone's TCB.

\section*{Acknowledgements}
We would like to thank Kevin Laeufer for providing helpful feedback throughout the project.

\bibliographystyle{unsrt}
\bibliography{main.bib}

\end{document}